# Estudio de los protocolos de Cero Conocimiento y Criptografía de curva elíptica y su implementación en ambientes de Tarjetas Inteligentes utilizando Java Card.

**Carlos Andres Agudelo Serna**

The University of Melbourne, Departamento de Ingeniería de Sistemas y Computación,
Melbourne, Australia

cagudelo@student.unimelb.edu.au

**Abstract**

This paper studies the problem of Zero Knowledge Protocol (ZKP) and elliptic curve cryptographic implementation on a computationally limited environment such as, the smart cards, using Java Card. Besides that, it is explained how the zero knowledge protocol was selected to implement it on a smart card and how the benchmarking was made in order to select this protocol. It is shown a theoretical development to implement the ZKP protocol using elliptic curve cryptography.

**Resumen**

Este artículo estudia la problemática de la implementación de protocolos de cero conocimiento y funciones de criptografía de curva elíptica en un ambiente limitado computacionalmente como lo son las tarjetas inteligentes utilizando la tecnología de Java Card. Adicionalmente se explica como fue la selección del protocolo de cero conocimiento a implementar en la tarjeta y el estudio de desempeño realizado para la selección del mismo. Se plantea un desarrollo teórico para la implementación del mismo protocolo seleccionado, utilizando criptografía de curva elíptica.

**Palabras claves:** Autenticación, Cero conocimiento, Criptografía, Curva elíptica, Java card, Tarjetas inteligentes

1. **INTRODUCCIÓN**

Actualmente con el surgimiento de nuevas y múltiples redes de comunicación, en particular de Internet, se han abierto nuevas posibilidades para el intercambio de la información, surgiendo nuevas amenazas a la seguridad de los datos que se pretenden transmitir, en este aspecto es donde la criptografía moderna juega un papel primordial, dado que esta se preocupa por garantizar principalmente la confidencialidad, autenticidad, integridad y no repudio de la información a través de algoritmos criptográficos encargados de cifrar dicha información para la transmisión de la misma sobre canales inseguros como es el caso de Internet.

Teniendo en cuenta lo anterior, la identificación basada en infraestructura de llave pública (PKI – Public Key Infrastructure) es una herramienta muy útil y fundamental en muchas de estas aplicaciones tales como transferencia de fondos electrónicos y sistemas en línea para prevenir el acceso no autorizado de usuarios inválidos. Es en estos escenarios donde la identificación juega un papel muy importante, dado que garantiza la autenticidad de la información que se procesa. Los esquemas de Identificación se pueden implementar como aplicaciones de protocolos basados en pruebas interactivas de cero conocimiento donde no se revela información que pueda comprometer el sistema.

La criptografía de curva elíptica proporciona una manera eficaz y eficiente de implementar este tipo de protocolos en ambientes limitados computacionalmente, dado que el tamaño de la llave empleada puede ser mucho mas pequeña lo que la hace susceptible de ser implementada en este tipo de plataformas como PDA's (Personal Digital Assistant), teléfonos celulares, tarjetas inteligentes, entre otros. Este tipo de criptografía ha tenido un crecimiento explosivo en los últimos años, a tal grado que es uno de los dos tipos de criptografía de llave pública mas utilizados en todo el mundo (En compañía de RSA – Rivest Shamir Adleman).

El objetivo de este articulo es estudiar como se podrían implementar protocolos de cero conocimiento y criptografía de curva elíptica en arquitecturas de bajos recursos computacionales como es el caso de las tarjetas inteligentes, utilizando la tecnología de Java Card y frameworks como el JCA (Java Cryptographic Arquitecture), el JCE (Java Cryptographic extensión) y el JCSP (Java Cryptographic Service Provider), el SDK (Software Development Kit) de la compañía Schlumberger y el Middleware proporcionado por la misma compañía.

El articulo esta organizado de la siguiente manera: en la sección 2, se describe básicamente los protocolos de cero conocimiento, sus propiedades y características. En la sección 3 se indica como fue el proceso de evaluación de desempeño y los criterios tomados en cuenta para seleccionar el protocolo mas apropiado a implementar sobre las tarjetas inteligentes. En la sección 4, se explica brevemente algunos conceptos básicos sobre criptografía de curva elíptica necesarios. En la sección 5, se describe los protocolos de la sección 2 utilizando curva elíptica. En la sección 6, se plantea el esquema de acuerdo a los resultados obtenidos en la sección 5. En la sección 7, se describe de una manera básica la arquitectura de Java Card y sus limitaciones. Finalmente, en la sección 8, se encuentran las conclusiones halladas en la realización de este artículo.

## 2. PROTOCOLOS DE CERO CONOCIMIENTO

Los protocolos de cero conocimiento se utilizan básicamente cuando una parte (El Demostrador – Prover) quiere probar a una contraparte (El Verificador – Verifier) el conocimiento de un secreto sin revelar dicho secreto en el proceso, de esta manera el verificador no obtiene información confidencial susceptible de ser comprometida en el sistema.

Una prueba interactiva usualmente toma la forma de un protocolo de desafío – respuesta, en la cual el *demostrador* y el *verificador* intercambian mensajes y el *verificador* acepta o rechaza al final del protocolo únicamente si todas las pruebas son aceptadas como validas.

Los protocolos de cero conocimiento deben tener las siguientes propiedades, especialmente en aplicaciones criptográficas: [5]

- **Completitud (*Completeness*).** El *verificador* siempre acepta la prueba si el hecho es verdadero y ambos el *demostrador* y el *verificador* siguen el protocolo.

- **Validez (*Soundness*).** El *verificador* siempre rechaza la prueba si el hecho es falso, en tanto, el *verificador* siga el protocolo.

- **Cero Conocimiento (*Zero Knowledge*).** El *verificador* no aprende nada sobre el hecho que esta siendo demostrado por su contraparte (excepto que es correcto y veraz). En una prueba de cero conocimiento, el *verificador* no puede demostrar o probar después el hecho a alguien más.

Los protocolos de cero conocimiento tienen las siguientes características:

- Cada acierto certifica más a un usuario y una falla lo desacredita del todo. Este proceso recibe el nombre de acreditación, donde un usuario es acreditado n veces si el protocolo ha sido exitoso $2^n$ veces.

- La prueba proporciona una acumulación de confianza, no de conocimiento (no conclusiva), es decir, es una prueba probabilística y no absoluta.

- Por cada iteración exitosa del protocolo se adquiere un mayor nivel de confianza.

- El nivel de seguridad es configurable pues depende del número de iteraciones o pruebas del protocolo.

- La probabilidad de engaño es de $1/2^n$, donde "n" es el número de pruebas o iteraciones realizadas.

- Por lo general las preguntas son de respuesta binaria por ejemplo, si/no, izquierda/derecha, arriba/abajo, verdadero/falso, entre otros.

Una iteración típica en una prueba de cero conocimiento consiste en un mensaje de "compromiso" que viene del *demostrador*, seguido por un reto proveniente del *verificador* y luego una respuesta al reto proveniente del *demostrador*. El protocolo puede repetirse varias veces (múltiples iteraciones). Basado en las respuestas del *demostrador* en todas las iteraciones, el *verificador* decide aceptar o rechazar la prueba, esto se observa en la figura 1, donde se muestra el funcionamiento general de un protocolo de cero conocimiento.

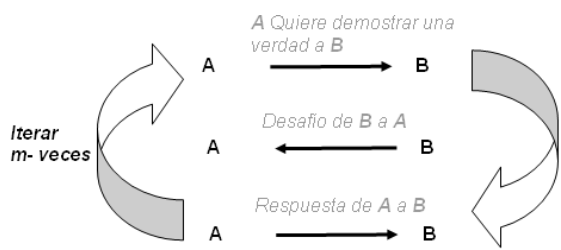

Figura 1. Funcionamiento general de un protocolo de cero conocimiento.

Un ejemplo clásico es el de la cueva de Ali Baba. [5].

### 2.1 Los Problemas Clásicos
Existen varios problemas clásicos que involucran protocolos de cero conocimiento. En este artículo se presentan dos de estos problemas, principalmente el problema de las raíces cuadradas y el problema del logaritmo discreto. []

### 2.2 El Problema de las Raíces Cuadradas
En el problema de las raíces cuadradas *el demostrador* quiere probar a través de un protocolo de cero conocimiento que conoce la raíz cuadrada de un numero determinado modulo un numero *n*, el cual es un número compuesto grande de factorización desconocida (por ejemplo un numero de 1024 bits producto de dos números primos). Es decir, se debe probar a través de un protocolo que satisfaga las tres propiedades de un protocolo de cero conocimiento (Completitud, validez y cero conocimiento) que *el demostrador* conoce *x* tal que:

$$x^2 = b \ (mod.\ n),$$ donde se conocen *b* y *n*.

Los protocolos estudiados y analizados que pertenecen a este grupo, es decir, que su seguridad esta basada en este problema son: el protocolo de residuos cuadráticos, el protocolo de Fiat-Shamir y el protocolo de Guillou-Quisquater. Estos protocolos son variaciones de este problema y tienen diferentes fases y configuraciones de acuerdo al número de mensajes intercambiados y los requerimientos computacionales. [7]

Básicamente el funcionamiento de un protocolo de cero conocimiento basado en el problema de las raíces cuadradas seria de la siguiente manera:

*El demostrador* genera un numero aleatorio *r* y calcula $s = r^2 \ mod.\ n$. Luego envía *s* al *verificador*, el cual genera un bit aleatorio *c* y de acuerdo al resultado pregunta al *demostrador* por *r* o $m = r.x$. *El verificador* verifica este valor y repite la prueba el número de veces necesario hasta quedar satisfecho con el resultado, de otra manera la prueba se considera inválida. La tabla 1 explica este proceso de manera resumida.

| Pasos | Descripción | Demostrador (P) | Verificador (V) |
|---|---|---|---|
| 0 | | b,n,x | b,n |
| 1 | El demostrador genera el numero aleatorio *r* | r | |
| 2 | P envía $s = r^2 \ mod.\ n$ a V | s | s |
| 3 | V genera un bit aleatorio   c={0, 1} | c | c |
| 4 | Si *c = 0* entonces P envía *r* a V | | Verifica $r^2=s$ |
| 5 | Si c = 1 entonces P envía $m = r.x$ | | Verifica $m^2=s.b$ |
| 6 | Los pasos 1-5 se repiten hasta que el Verificador este convencido que el Demostrador debe conocer *x* (con una probabilidad $1-2^{-k}$, para k iteraciones) | | |

Tabla 1. Protocolo de cero conocimiento basado en el problema de raíces cuadradas.

### 2.3 El Problema del Logaritmo Discreto
En el problema del logaritmo discreto *el demostrador* quiere probar a través de un protocolo de cero conocimiento que conoce el logaritmo discreto de un numero determinado modulo un número *n*. Es decir, dado *n* (modulo), el generador *g* para un campo finito *Fn* y $b \in Fn$, se debe demostrar a través de un protocolo que cumpla las tres propiedades de cero conocimiento que *el demostrador* conoce *x* tal que:

$$g^x = b \ (mod.\ n).$$

El problema del logaritmo discreto es uno de los problemas computacionales más difíciles de resolver actualmente. Por lo tanto, se utiliza en protocolos de cero conocimiento donde se desea demostrar el conocimiento de un determinado secreto sin revelarlo en el proceso.

El protocolo estudiado y analizado que basa su seguridad en este problema es el protocolo de Schnorr, el cual fue escogido para la implementación sobre la tarjeta inteligente, ya que reduce los cálculos en tiempo real para *el demostrador* a una multiplicación modulo un primo *q*. Esto es muy útil cuando *el demostrador* tiene poder de cómputo muy limitado como es el caso de las tarjetas inteligentes y a la eficiencia que fue determinada en las pruebas de desempeño realizadas. [7], []. Las pruebas realizadas se detallan más adelante en la sección 3.

El algoritmo de cero conocimiento seleccionado (Schnorr) basa su eficiencia computacional en el uso de un subgrupo de orden *q* del grupo multiplicativo de enteros modulo *p*. Donde $q|(p-1)$ (p-1 es divisible por q). Esto reduce el número requerido de bits transmitidos. Este protocolo fue diseñado para realizarse únicamente en tres pasos y requiere un bajo ancho de banda en comunicaciones comparado con Fiat-Shamir, Guillou-Quisquater y el protocolo basado en residuos cuadráticos.

El funcionamiento de un protocolo de cero conocimiento basado en el problema del logaritmo discreto es de la siguiente manera:

*El demostrador* genera un numero aleatorio *r* y calcula $h = g^x$ *mod. n*. Posteriormente envía *h* al *verificador*, luego este genera un bit aleatorio *c* y lo envía al *demostrador*, si el bit es cero *el demostrador* envía *r* al *verificador* y este verifica que $g^r = h$. Si el bit es uno, *el demostrador* envía $m = x + r$ al *verificador* y este verifica que $g^m = b.h$. Estos pasos se repiten hasta que *el verificador* este convencido que *el demostrador* debe conocer *x* con una probabilidad de $1-2^{-k}$, donde *k* es el numero de iteraciones del protocolo. La tabla 2 resume el funcionamiento de este protocolo.

| Pasos | Descripción | Demostrador (P) | Verificador (V) |
|---|---|---|---|
| 0 | | g,b,n,x | g,b,n |
| 1 | El demostrador genera el numero aleatorio *r* | r | |
| 2 | P envía $h = g^r$ *mod. n* a V | h | h |
| 3 | V genera un bit aleatorio    c={0, 1} | c | c |
| 4 | Si *c = 0* entonces P envía *r* a V | | Verifica $g^r=h$ |
| 5 | Si c = 1 entonces P envía $m = x+ r$ | | Verifica $g^m=b.h$ |
| 6 | Los pasos 1-5 se repiten hasta que el Verificador este convencido que el Demostrador debe conocer *x* (con una probabilidad $1-2^{-k}$, para k iteraciones) | | |

Tabla 2. Protocolo de cero conocimiento basado en el problema del logaritmo discreto.

## 3. PRUEBAS REALIZADAS PARA SELECCIONAR EL PROTOCOLO DE CERO CONOCIMIENTO

En esta sección se realizaron una serie de pruebas sencillas para determinar el desempeño de los protocolos de cero conocimiento mencionados anteriormente sobre un ambiente de tarjetas inteligentes.

Los protocolos de Residuos cuadráticos, Fiat – Shamir, Guillou – Quisquater y Schnorr [] proveen soluciones al problema de la identificación. Cada uno tiene ventajas y desventajas relativas de acuerdo a los criterios de evaluación que se utilicen y a las aplicaciones especificas que se deseen implementar. Para comparar los protocolos se deben escoger un conjunto típico de parámetros seleccionados para cada uno de los protocolos y teniendo en cuenta los niveles de seguridad que se desean implementar.

Para las comparaciones realizadas se tuvieron en cuenta los siguientes criterios:

- Comunicaciones o Ancho de banda: Indica los números de mensajes intercambiados, esto traducido en el número total de bits transmitidos.

- Cálculos: Indica el número de multiplicaciones modulares requeridas para ambos *el demostrador* (Prover) y *el verificador* (Verifier), teniendo en cuenta los cálculos on-line y off-line (como es el caso de Schnorr).

- Memoria: Indica los requisitos de almacenamiento necesarios en la tarjeta para guardar las llaves secretas y efectuar los cálculos necesarios.

- Tiempo: Indica el tiempo en milisegundos que demora la ejecución del protocolo de cero conocimiento de acuerdo a los parámetros de configuración.

Las pruebas realizadas se hicieron sobre una tarjeta Cyberflex Access e-gate 32K, utilizando la versión 2.1.1 del API de Java Card y el kit de desarrollo Cyberflex Access SDK versión 4.5, la comunicación entre la tarjeta y el PC se hizo a través de un lector e-gate token connector USB. El computador utilizado para las pruebas es un Pentium IV de 2.8 Ghz con un 1 GB de memoria RAM, bus de 800 MHZ y 512 KB de memoria Cache L1.

Es importante resaltar que los resultados se tomaron teniendo en cuenta el demostrador (Prover) que para nuestro caso es la tarjeta Inteligente, ya que el papel del verificador lo desempeña un computador de escritorio con la configuración especificada anteriormente y no es relevante para el análisis que se pretende desarrollar en este articulo.

La configuración de los parámetros para la realización de las pruebas realizadas fue la siguiente:

- Multiplicaciones modulares con un modulo n de 256 bits

- El numero de iteraciones *t* fue definido con un valor de 100

- La función de hashing empleada para los cálculos que lo requerían fue MD5 (128 bits).

- La longitud *k* de los vectores de números enteros de inicialización fue 3. para los protocolos de Fiat – Shamir y Guillou – Quisquater.

Tomando el primer criterio de comparación, el de comunicaciones o ancho de banda, se obtuvo el siguiente resultado que se puede apreciar en la figura 2.

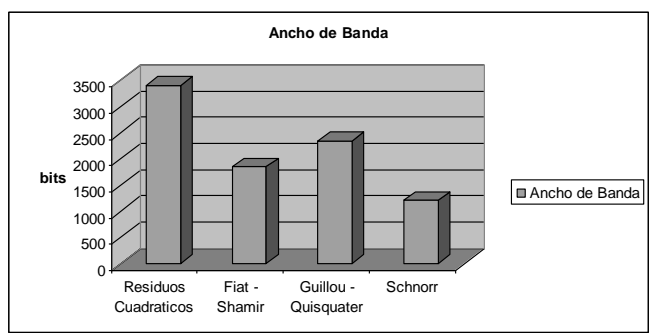

Figura 2. Consumo del ancho de banda de los protocolos de cero conocimiento

Según la grafica anterior los protocolos de cero conocimiento de mayor consumo de ancho de banda son el de residuos cuadráticos y el de Guillou – Quisquater, según lo analizado la razón puede ser básicamente porque se debe enviar la Identidad ($I_A$) al verificador como parte del mensaje de *commitment* o compromiso lo que podría generar un poco de *overhead* en la transmisión entre *el verificador* y *el demostrador*, es decir, entre el computador y la tarjeta inteligente. Sin embargo en el protocolo de GQ se pueden reducir los parámetros de memoria (k) y los parámetros de la transmisión (t). El protocolo que menos consume ancho de banda es el de Schnorr debido a que la información que envía el Probador es muy pequeña comparado con los otros algoritmos, dado que el numero de iteraciones (t) en el protocolo de Schnorr es mucho mas reducido comparado con Fiat-Shamir y residuos cuadráticos.

El numero de cálculos realizados por *el demostrador*, es decir, la tarjeta inteligente se resumen en la figura 3 donde se observa claramente que el protocolo que menos realiza multiplicaciones modulares es el de Schnorr, este protocolo tiene la ventaja que solo requiere una única multiplicación modular en línea por *el demostrador* (*prover*), obviamente se requiere muchos mas cálculos del lado del *verificador* comparado con otros protocolos como Fiat-Shamir, GQ o residuos cuadráticos, esto le agrega un poco de complejidad en el lado del *verificador*, es decir, el computador pues debe realizar mas cálculos y verificaciones.

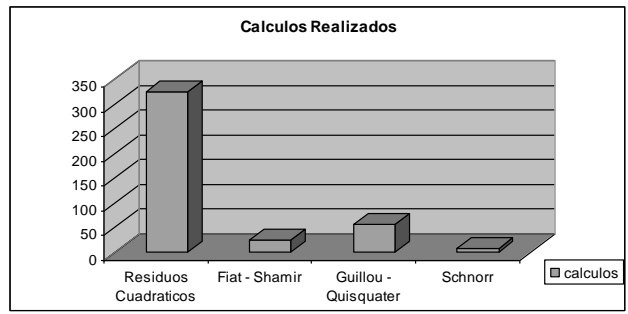

Figura 3. Cálculos computacionales efectuados por los protocolos de cero conocimiento

En la figura 3 se observa que el numero de cálculos realizados por el protocolo de residuos cuadráticos es bastante alto comparado con el resto de protocolos, esto se debe básicamente a que se deben calcular todos los números relativamente primos al modulo *n* y sus respectivos residuos cuadráticos, lo cual es una carga muy pesada para arquitecturas de recursos limitados como las tarjetas Inteligentes produciendo además una carga adicional en tiempo y memoria necesaria.

El siguiente criterio evaluado es el consumo de memoria que se ve reflejado en la figura 4, como era de esperarse el protocolo de residuos cuadráticos es el protocolo que mayor consumo de memoria tiene, debido a los múltiples cálculos realizados, al envío de mensajes entre *demostrador* y *verificador* y principalmente los requisitos de almacenamiento necesario para las llaves empleadas en el proceso.

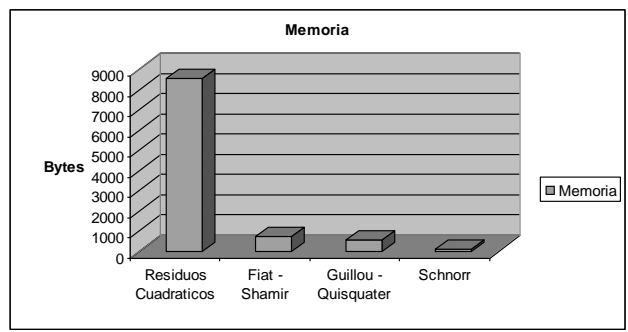

Figura 4. Consumo de memoria de los protocolos de cero conocimiento

Por otro lado se observa que el consumo de memoria del protocolo Schnorr es mínimo, pues como se explico anteriormente los cálculos y el consumo de memoria del lado del *verificador* (tarjeta inteligente) son pequeños, lo cual contrasta con la gran cantidad de cálculos y consumo de memoria del protocolo del lado del v*erificador* (servidor, PC, entre otros).

Finalmente, el último criterio evaluado fue el consumo de tiempo de los diferentes protocolos de cero conocimiento planteados. Se observa en la figura 5 que el mayor consumo de tiempo pertenece al protocolo de residuos cuadráticos, debido a los múltiples cálculos y requerimientos que necesita para su ejecución.

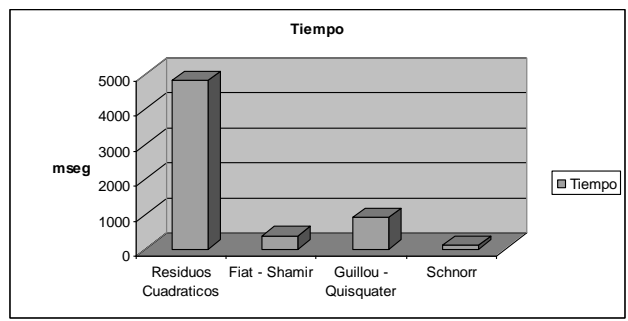

Figura 5. Consumo de tiempo de los protocolos de cero conocimiento

El protocolo más eficiente en cuanto a tiempo es de nuevo el protocolo de Schnorr, dado la sencillez de la multiplicación modular del lado del *demostrador*, pero las operaciones son un poco más complejas al igual que los requerimientos de memoria del lado del *verificador* como se mencionó anteriormente.

Cabe mencionar que el protocolo de Guillou-Quisquater aunque requiere menos consumo de memoria toma mas tiempo en realizar las operaciones pertinentes, es decir, multiplicaciones modulares, dado el grado de complejidad y la cantidad de operaciones realizadas en comparación con el protocolo de Schnorr, donde solo se realiza una sola operación modular.

En la tabla 2 se encuentran los valores hallados durante la comparación y evaluación de desempeño realizada sobre el ambiente de tarjeta inteligente descrito anteriormente de los protocolos de cero conocimiento: Residuos Cuadráticos, Fiat-Shamir, Guillou-Quisquater y Schnorr.

| Protocolos de Cero Conocimiento | | | | |
|---|---|---|---|---|
| Protocolo<br>Criterio | **Residuos Cuad.** | **Fiat – Shamir** | **Guillou –Quisq** | **Schnorr** |
| Ancho de Banda (bits) | 3400 | 1850 | 2350 | 1207 |
| Cálculos | 325 | 26 | 57 | 8 |
| Memoria (bytes) | 8526 | 725 | 578 | 137 |
| Tiempo (mseg) | 4825 | 387 | 900 | 123 |

Tabla 3. Criterios de evaluación utilizados para los protocolos de cero conocimiento

Los datos mostrados en esta tabla corresponden a los gráficos de los diferentes criterios de evaluación tomados en cuenta: ancho de banda, cálculos, memoria y tiempo.

Observando los resultados obtenidos en la evaluación de desempeño de los diferentes protocolos de cero conocimiento analizados se sugiere que el protocolo óptimo a implementar sobre un ambiente de tarjetas inteligentes para un esquema de identificación o firmas digitales basados en cero conocimiento seria el protocolo de Schnorr, teniendo en cuenta que se debería implementar del lado del servidor o equipo de computo encargado del papel de *verificador* un mayor numero de operaciones y cálculos en comparación con los cálculos requeridos por el *verificador* en protocolos como Fiat-Shamir y Guillou-Quisquater.

## 4. CRIPTOGRAFÍA DE CURVA ELÍPTICA (ELLIPTIC CURVE CRIPTOGRAPHY – ECC)

La criptografía de curva elíptica, propuesta por Miller y Koblitz en 1985, se fundamenta básicamente sobre matemáticas de curva elíptica, las cuales son un poco más complejas que las empleadas por otros criptosistemas de llave pública como RSA.

Una curva elíptica $E$ definida sobre un campo $F$ puede definirse como un conjunto de elementos (x,y) que satisfacen la ecuación:

$$y^2 + a_1 xy + a_3 y = x^3 + a_2 x^2 + a_4 x + a_6$$

conjuntamente con un punto especial $O$ que no tiene coordenadas y es llamado punto al infinito, esta representación se conoce como la forma "Weierstrass" de una curva elíptica. Cuando el campo $F$ es un conjunto finito de números enteros (campo finito) la ecuación anterior queda reducida de la siguiente manera [2]:

$$y^2 = x^3 + a_4 x + a_6$$

Los campos finitos pueden describirse como un conjunto cerrado de elementos que cumplen con ciertas propiedades [2]:

- Existe una operación de adición de elementos y el campo contiene al elemento neutro aditivo.

- Existe una operación de multiplicación de elementos y él contiene el elemento neutro multiplicativo.
- Cada elemento en un campo posee un inverso aditivo y multiplicativo.

Para efectos criptográficos, dos campos son de gran interés al momento de definir curvas elípticas [8]:

- Campos finitos primos ($F_p$): estos campos se encuentran compuestos por un número primo p de elementos. Los elementos de este campo es el conjunto de enteros módulo p y la aritmética en este tipo de campos es de tipo modular.

- Campos finitos binarios ($F_{2^m}$): estos campos poseen $2^m$ elementos para un m dado, conocido como el grado del campo. Los elementos de este campo son las cadenas binarias de tamaño m y la aritmética en este tipo de campos es implementada en términos de operaciones de bits.

Las curvas recomendadas por seguridad y rendimiento son las basadas en campos $F_{2^m}$ ya que éstas permiten una mayor velocidad de procesamiento dado que sus elementos principales son cadenas binarias que favorecen las operaciones lógicas AND y XOR, así como también las operaciones de aritmética modular.

Un aspecto importante dentro de la criptografía de curva elíptica es el de la adición de puntos. Los puntos sobre una curva forman un grupo abeliano[1] bajo la operación de adición. Este grupo abeliano cumple con las leyes básicas de los campos sobre la operación adición, es decir posee un elemento neutro aditivo, cada elemento posee su inverso aditivo. Teniendo en cuenta estas leyes y considerando un campo de característica 2, es decir $F_{2^m}$, podemos considerar la suma de dos puntos P y Q sobre una curva, como se muestra en la figura 6:

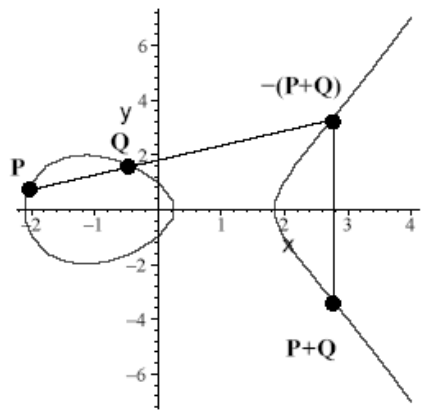

Figura 6. Adición de dos puntos sobre la curva $y^2 = x^3 - 4x + 1$ []

La criptografía de curva elíptica basa su fortaleza, al igual que RSA, en un problema matemático, el problema del logaritmo discreto de curva elíptica, ECDLP (Elliptic Curve Discrete Logarithm Problem). Este problema se resume de la siguiente forma [9]: asuma sobre una curva *E* el producto *x . P* esto representa el punto *P* adicionado a él mismo *x* veces. Suponga que *Q* es un múltiplo de *P*, tal que: *Q = x . P* para algún *x*. Entonces el problema del algoritmo discreto de curva elíptica es determinar *x* dado *P* y *Q*.

Actualmente no se conoce ningún mecanismo que pueda resolver este problema eficientemente, y los ataque más conocidos aun tardan tiempos exponenciales. Sin embargo, existe un conjunto de curvas recomendadas por el NIST [8], las cuales son menos vulnerables a ataques y por lo tanto mucho mas seguras. La tabla 4 muestra los tamaños recomendados de los campos empleados para definir las curvas, equivalentes entre primos y binarios. En [8] se encuentra de forma más específica los parámetros establecidos para la definición de las curvas pertenecientes a cada uno de estos campos.

| Campo Primo | Campo Binario |
|---|---|
| ‖ p ‖ = 192 | m = 163 |
| ‖ p ‖ = 224 | m = 233 |

---

[1] Se denomina grupo conmutativo o abeliano a aquel grupo que verifica la propiedad conmutativa, es decir que para todo x, y de **X**: :x*y = y*x

| ‖ p ‖ = 256 | m = 283 |
| ‖ p ‖ = 384 | m = 409 |
| ‖ p ‖ = 521 | m = 571 |

Tabla 4. Tamaños de campos recomendados por NIST [8]

En la anterior tabla, el valor ‖ p ‖ representa la longitud de la expansión binaria del entero p.

La criptografía de curva elíptica se encuentra especificada principalmente por los estándares ANSI X9 e IEEE P163, los cuales incluyen el algoritmo de cifrado ECIES (Elliptic Curve Integrated Encryption Scheme), el de firma digital ECDSA (Elliptic Curve Digital Signature Algorithm) e el de intercambio de llaves ECKAS (Elliptic Curve Key Agreement Scheme).

La criptografía de curva elíptica requiere un tamaño de llave mucho menor al de RSA, dado que el problema del logaritmo discreto sobre curvas elípticas es mucho más complejo de resolver y requiere tiempo exponencial, en contraste al tiempo sub-exponencial requerido sobre campos ordinarios $Z_n$. Esto se ilustra en la tabla 5.

| Nivel de seguridad | Cifrado de Bloques | ECC $\|q\|_2$ | RSA $\|n\|_2$ |
|---|---|---|---|
| 80 | SKIPJACK | 160 | 1024 |
| 112 | Triple-DES | 224 | 2048 |
| 128 | AES small | 256 | 3072 |
| 192 | AES médium | 384 | 8192 |
| 256 | AES large | 512 | 15360 |

Tabla 5. Comparación del tamaño de la llave ECC vs. RSA []

Como se observa en la tabla ECC presenta una diferencia radical en cuanto a seguridad, es decir, un modulo de 300 bits en ECC es radicalmente más seguro que un modulo de 2048 bits en RSA. El tiempo necesario para resolver una instancia del problema del logaritmo discreto sobre curva elíptica es totalmente exponencial, en comparación con otros algoritmos basados en problemas como el problema de la factorización de enteros o el problema de logaritmo discreto como RSA y DSA (Digital Signature Algorithm) que cuyo tiempo es sub-exponencial, esto se observa claramente en la figura 7.

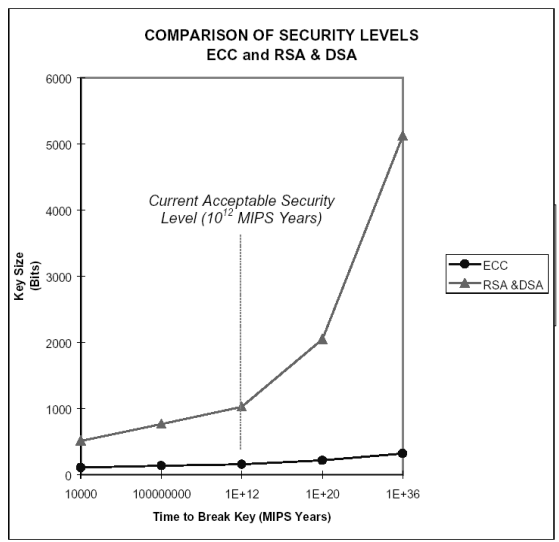

Figura 7. Comparación de tiempos para resolver ECC, RSA y DSA. []

Los tiempos se calculan en la unidad MIPSY (Millions of Instructions per Second – Years), normalmente en una prueba de desempeño se considera $10^{12}$ MIPSY como un nivel de seguridad razonable con los recursos computacionales actuales. En la figura 7, los tiempos necesarios para romper RSA y DSA se agrupan juntos dado que los algoritmos para resolver el problema de factorización de enteros sobre el cual se basa RSA y el problema del Logaritmo Discreto sobre el cual se basa DSA tienen aproximadamente los mismos tiempos de ejecución asintóticos. []

# 5. PROTOCOLOS DE CERO CONOCIMIENTO UTILIZANDO CRIPTOGRAFÍA DE CURVA ELÍPTICA

En esta sección se muestra como utilizar criptografía de curva elíptica en protocolos de cero conocimiento, básicamente se toman los problemas clásicos planteados en la sección 2 (2.2, 2.3) y se indican como implementarlos sobre curvas elípticas. Adicionalmente se indica por que funciona solo con el problema del logaritmo discreto y no con el problema de las raíces cuadradas. []

## 5.1 El Problema de las Raíces Cuadradas Sobre Curvas Elípticas

Sobre una curva elíptica $E$ definida sobre un campo $F_n$, $E/F_n$ (donde $n$ es un numero compuesto), y $B \in E/F_n$, El *demostrador* desea probar en un protocolo de cero conocimiento que conoce $A \in E/F_n$, tal que $2A = B$, es decir, $A+A = B$, dado que $n$ es un numero compuesto (producto de dos números primos diferentes), resolver este problema es computacionalmente muy difícil.

El funcionamiento del protocolo seria el siguiente:

*El demostrador* desea probar que conoce $A$, tal que $2A = A+A = B$. Primero, *el demostrador* genera un punto aleatorio $R \in E/F_n$ y calcula $S = 2R$ y envía S al *verificador* y este calcula un bit aleatorio $c$, si el bit aleatorio $c$ es cero, entonces *el demostrador* debe enviar $R$ al *verificador* y este verifica que $2R = S$. Si el bit aleatorio $c$ es uno, entonces *el demostrador* debe enviar $M=R+A$ al *verificador* y este verifica que $2M = S+B$. Los pasos se repiten hasta que *el verificador* este convencido de la validez de la prueba. La tabla 6 ilustra el funcionamiento de este protocolo.

| Pasos | Descripción | Demostrador (P) | Verificador (V) |
|---|---|---|---|
| 0 | | A,B | B |
| 1 | El demostrador genera el punto aleatorio $R$ | R | |
| 2 | P envía S $=2R$ a V | S | S |
| 3 | V genera un bit aleatorio $c=\{0, 1\}$ | c | c |
| 4 | Si $c = 0$ entonces P envía $R$ a V | | Verifica $2R = S$ |
| 5 | Si c = 1 entonces P envía $M = R+A$ | | Verifica $2M= S+B$ |
| 6 | Los pasos 1-5 se repiten hasta que el Verificador este convencido que el Demostrador debe conocer $x$ (con una probabilidad $1-2^{-k}$, para k iteraciones) | | |

Tabla 6. Protocolo de cero conocimiento basado en el problema de raíces cuadradas sobre curvas elípticas.

Se observa que *el verificador* en el paso 2 del protocolo descrito anteriormente puede resolver $R$ al factorizar $n$ en tiempo sub-exponencial, entonces podría engañar al *demostrador* en el paso 3 al fijar el bit aleatorio $c$ en uno y así obligar al *demostrador* a enviarle $M = R+A$, una vez *el verificador* obtiene $M$, puede conocer el secreto $A$ en tiempo sub-exponencial, dado que $A=M-R$. Esta debilidad en el protocolo se debe a que el esquema no está basado totalmente en el problema del logaritmo discreto sobre curvas elípticas y por lo tanto estas partes vulnerables pueden ser atacadas en tiempo sub-exponencial.

## 5.2 El Problema del Logaritmo Discreto Sobre Curvas Elípticas

Sobre una curva elíptica $E$ definida sobre un campo $F_n$, $G \in E/F_n$ (donde $G$ es un generador), y $B = m \cdot G \in E/F_n$, El *demostrador* desea probar en un protocolo de cero conocimiento que conoce $m$, tal que $m \cdot G = B$.

El funcionamiento del protocolo seria el siguiente:

*El demostrador* desea probar que conoce $m$, tal que $m \cdot G = B$, donde B es publico, entonces genera un número aleatorio $r \in F_n$ y calcula $A = r \cdot G$ y envía A al *verificador*. El *verificador* genera un bit aleatorio $c$ y lo envía al *demostrador*. Si el bit $c$ es cero, entonces *el demostrador* envía $r$ al *verificador* y este verifica que $r \cdot G = A$. Si el bit $c$ es uno, entonces *el demostrador* envía $x = r + m$ al *verificador* y este verifica que $x \cdot G = A+B$. La repetición de estos pasos incrementa exponencialmente la confianza (la probabilidad de engaño disminuye) del *verificador* hacia el *demostrador* y que este conoce el secreto $m$. La tabla 7 resume el funcionamiento de este protocolo.

| Pasos | Descripción | Demostrador (P) | Verificador (V) |
|---|---|---|---|
| 0 | | G,B,m | G,B |
| 1 | El demostrador genera el numero aleatorio $r$ | R | |
| 2 | P envía $A = r \cdot G$ a V | A | A |
| 3 | V genera un bit aleatorio $c=\{0, 1\}$ | c | c |
| 4 | Si $c = 0$ entonces P envía $r$ a V | | Verifica $r \cdot G = A$ |

| | | |
|---|---|---|
| 5 | Si c = 1 entonces P envía x = r + m | Verifica $x \cdot G = A + B$ |
| 6 | Los pasos 1-5 se repiten hasta que el Verificador este convencido que el Demostrador debe conocer *m* (con una probabilidad $1-2^{-k}$, para k iteraciones) | |

Tabla 7. Protocolo de cero conocimiento basado en el problema del logaritmo discreto sobre curva elíptica.

Como se observa el problema del logaritmo discreto sobre curva elíptica proporciona mayor seguridad al esquema del protocolo de cero conocimiento que el proporcionado en el esquema tradicional planteado en la sección 2.3 el cual está basado en el problema del logaritmo discreto sobre grupos multiplicativos *Zn*. Esto se debe principalmente a que resolver el problema del logaritmo discreto sobre curvas elípticas requiere tiempo exponencial, mientras que resolver el problema del logaritmo discreto sobre grupos multiplicativos *Zn* requiere tiempo sub-exponencial. []

# 6. ESQUEMA PROPUESTO DE IDENTIFICACIÓN BASADO EN EL PROBLEMA DEL LOGARITMO DISCRETO SOBRE CURVAS ELÍPTICAS

En esta sección se describe el esquema planteado, el cual está basado en el protocolo de cero conocimiento de Schnorr [], que basa su seguridad en el problema del logaritmo discreto sobre grupos multiplicativos *Zn*, lo que se pretende mostrar es como se podría migrar este esquema de identificación hacia criptografía de curva elíptica a través del problema del logaritmo discreto sobre curvas elípticas. La principal motivación para utilizar curvas elípticas sobre un campo finito es básicamente las ventajas que representa y la posibilidad de ser implantando sobre arquitecturas de recursos computacionales limitados como las tarjetas inteligentes, dado que el tamaño de la llave puede ser mucho más pequeño que en otros esquemas, debido a que el tiempo requerido para romper este algoritmo es exponencial y la seguridad es mayor en este esquema si se escogen cuidadosamente curvas elípticas especiales [7], las cuales garantizan una mayor seguridad y fortaleza criptográfica.

El algoritmo planteado es una reforma del algoritmo original de Schnorr [], esta propuesta presenta un esquema de identificación de tres pasos. El esquema hereda casi todos los beneficios del esquema original de identificación de Schnorr, tales como: el número reducido de interacciones, un tamaño de memoria reducido, velocidad de procesamiento y una comunicación reducida.

Como se dijo anteriormente la implementación de Schnorr sobre curvas elípticas permite incrementar el nivel de seguridad radicalmente debido al hecho que resolver el problema del logaritmo discreto sobre curvas elípticas requiere tiempo exponencial, en contraste al tiempo sub-exponencial que requiere el logaritmo discreto sobre el anillo *Zn*.

El esquema de identificación propuesto está basado en dos partes:

- La generación de la Llave (Selección de parámetros del dominio de la curva y generación de la llave)
- El protocolo Interactivo de tres mensajes entre *A* (*Demostrador*) y *B* (*Verificador*)

A continuación se explica cada una de estas partes:

**6.1 Generación de la Llave**
Primero se escogen los parámetros del dominio de la curva:

- Un tamaño de campo *q*, donde *q* es una potencia prima *q=p* un primo impar o *q=$2^m$*

- Se escogen dos elementos del campo *a,b* $\in$ *$F_q$* los cuales definen la ecuación de la curva elíptica *E* sobre el campo *$F_q$*, es decir, $y^2 = x^3 + ax + b$ en el caso *p > 3*, donde $4a^3 + 27b^2 \neq 0$.

- Se escogen dos elementos del campo *$x_p$, $y_p$*, los cuales definen un punto finito *P = ($x_p$, $y_p$)* de orden primo en *E($F_q$)*. El punto finito *P $\neq$ O*, donde *O* denota el punto al infinito.

- Se escoge el orden *n* del punto *P*.

La operación de la generación de la llave es el siguiente:

- Se selecciona un parámetro de seguridad *t*, el cual debe ser un entero positivo, por ejemplo *t $\geq$ 20*.

- Se selecciona *$P_1$,$P_2$* de orden *n* en el grupo *E ($F_q$)*

- Se seleccionan los enteros aleatorios $d_1$, $d_2$ del intervalo *[1,n-1]*. El par *($d_1$, $d_2$)* es la llave secreta.
- Se calculan los puntos $Q_1$ y $Q_2$ sobre *E,* tal que
- $Q_1 = -d_1 \bullet P_1$   $Q_2 = -d_2 \bullet P_2$
- Se calcula el punto *V* , tal que $V = Q_1 + Q_2$

**6.2 Protocolo Interactivo Entre *A* y *B***

El protocolo interactivo de tres pasos entre *A* (*Demostrador*) y *B* (*Verificador*) es el siguiente:

- *A* selecciona los números aleatorios $r_1$, $r_2 \in [1, n\text{-}1]$ y calcula el punto *Q* sobre *E*, tal que:
  $Q = r_1P_1 + r_2P_2 = (x_Q, y_Q)$  y  envía *Q* a *B*.
- *B* envía un número aleatorio $e \in Z_2^t$ a *A*.
- *A* envía a *B* el par ($y_1$,$y_2$), tal que: $y_1 = r_1 + ed_1$, $y_2 = r_2 + ed_2$
- *B* calcula el punto *W* sobre *E*.  $W = y_1 \bullet P_1 + y_2 \bullet P_2 + e \bullet V = (x_W, y_W)$ y verifica que: $x_Q = x_W$, donde $x_Q$ es la coordenada *x* de *Q* y $x_W$ es la coordenada *x* de *W*.
- *B* acepta la prueba como valida si y solo si $x_Q = x_W$, de otra manera la rechaza.

**6.3 Consideraciones de Seguridad del Esquema[]**

El tamaño de la llave puede ser mucho más pequeño que en otros esquemas de identificación, dado que, se necesitan ataques en tiempo exponencial para romper el criptosistema. La llave puede ser de 160 bits. Además si la curva es escogida cuidadosamente, el problema del logaritmo discreto sobre curvas elípticas podría ser intratable así se pueda factorizar y se rompa el logaritmo discreto del grupo multiplicativo.

Para evitar los ataques de Pollar-rho y Pohling-Hellman es necesario que el numero de Puntos racionales de Fq sobre *E #E(Fq)* sea divisible por un numero primo *n* grande.

Para evitar el ataque de reducción de MOV (Menezes, Okamoto, Vanstone), las curvas deben ser no-supersingulares, es decir, *p* no debe dividir (*q+1 - #E(Fq)*)

Para evitar el ataque de Semaev, Smart sobre curvas anómalas Fq , la curva no debe ser anómala en Fq, es decir, *#E(Fq) ≠ q.*

Una manera de evitar estos ataques y otros similares es seleccionar una curva elíptica *E* aleatoriamente sujeta a la condición que *#E(Fq)* sea divisible por un numero primo *n* grande, de esta manera la probabilidad que una curva aleatoria sucumba ante estos ataques es muy pequeña.

## 7. EL ENTORNO JAVA CARD Y LAS TARJETAS INTELIGENTES

Una tarjeta inteligente (Smart Card) es un dispositivo del tamaño de una tarjeta de crédito, el cual almacena y procesa información mediante un circuito de silicio embebido en el plástico de la tarjeta. [9]

Las tarjetas magnéticas (Magnetic Stripe Cards) utilizadas en cajeros automáticos y como tarjetas de crédito, son antecesoras de las tarjetas inteligentes. Las mismas almacenan información en una banda magnética de tres pistas que llevan adherida sobre la superficie. [9]

Las tarjetas inteligentes a diferencia de las tarjetas de banda magnética poseen todas las funciones necesarias e información para el procesamiento de datos, por ello no requieren acceso a bases de datos remotas durante una transacción. [9]

Las Smart Cards se pueden clasificar en dos tipos: [9]

- Memory Cards (también llamadas Low-End): Tienen circuitos de memoria que permiten almacenar datos. Estas tarjetas utilizan cierta lógica de seguridad, a nivel del hardware, para controlar el acceso a la información.
- Microprocessor Cards (también llamadas High-End): Tienen un microprocesador que les brinda una limitada capacidad de procesamiento de datos. Tiene capacidad de lectura, escritura y procesamiento.

Las Smart Cards contienen tres tipos de memoria: ROM, EEPROM y RAM. Normalmente no contienen fuente de poder, display o teclado. Interactúa con el mundo exterior utilizando una interfaz serial con ocho puntos de contacto como se muestra en la figura 8 las dimensiones y ubicación de los mismos están especificados en el estándar ISO 7816 (parte 2).

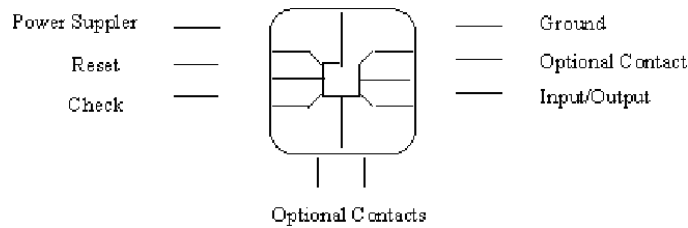

Figura 8. Arquitectura física de una tarjeta inteligente

Son utilizadas en conjunto con un Card Acceptance Device (CAD o Reader), que puede, aunque no necesariamente, estar conectado a una computadora (figura 9). Este dispositivo tiene como objetivo el proveer la fuente de poder e interfaz de comunicación para las tarjetas. La interacción entre la tarjeta y el CAD puede hacerse mediante un punto de contacto (la tarjeta debe ser insertada en el CAD) o sin contacto (utilizando radiofrecuencia).

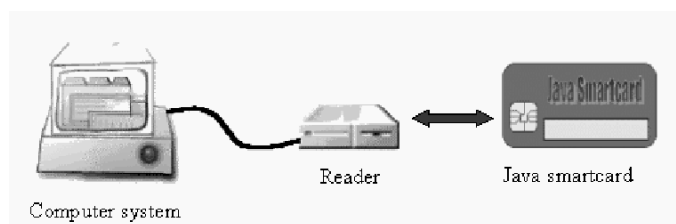

Figura 9. Interacción entre la tarjeta y el CAD

Dentro de la categoría de tarjetas inteligentes con microprocesador se encuentran las llamadas Java Cards o Java Smart Cards. Una Java Card es una tarjeta inteligente capaz de ejecutar programas desarrollados en Java. Originalmente, las aplicaciones para Smart Cards eran escritas en lenguajes específicos de los proveedores de Smart Cards (normalmente el ensamblador del microprocesador utilizado, o eventualmente C). La primera Java Card en salir al mercado fue producida por Schlumberger[2], aún antes que Sun[3] fijara el estándar.

En pocas palabras, una Java Card es una tarjeta con microprocesador que puede ejecutar programas (llamados applets) escritos en un subconjunto del lenguaje Java.

La arquitectura básica de una Java Card consiste de Applets, Java Card API, Java Card Virtual Machine (JCVM) / Java Card Runtime Environment (JCRE) y el sistema operativo nativo de la tarjeta (Figura 10).

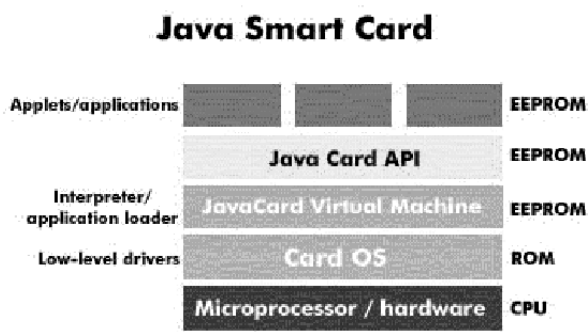

Figura 10. Arquitectura básica de Java Card.

---

[2] Empresa petrolera líder en el Mercado que ha patrocinado el apoyo de la tecnología, actualmente el desarrollo y soporte de aplicaciones de tarjetas inteligentes es manejado por la empresa Axalto.
[3] Sun Microsystems empresa creadora del lenguaje de programación Java.

La máquina virtual corre sobre el sistema operativo de la tarjeta, y se puede ver como una abstracción del mismo. Sobre la máquina virtual se encuentra el Java Card Framework, que es un conjunto de clases necesarias para el funcionamiento del JCRE y Java Card API, así como las clases utilitarias comerciales o extensiones propietarias del fabricante de la tarjeta. Finalmente, haciendo uso de este ambiente se encuentran los applets o aplicaciones de la tarjeta inteligente.

Existen múltiples aplicaciones y usos de las tarjetas inteligentes en la actualidad. Algunos ejemplos típicos se citan a continuación: [9]

- *Electronic Purse o Electronic Wallet (ePurse y eWallet)*
- *Transacciones Seguras*
- *Identificación Digital / Firma Digital*
- *Programas de Lealtad*
- *Sistemas de Prepago*
- *Health Cards*

La tecnología Java Card combina parte del lenguaje de programación Java con un entorno de ejecución optimizado para las tarjetas inteligentes y similares. El objetivo de la tecnología Java Card es llevar los beneficios del desarrollo de software en Java a las tarjetas inteligentes. [10] Algunas características relevantes del API de Java Card en comparación con el API de Java estándar se muestran en la siguiente tabla.

| Tecnología Java Card ||
|---|---|
| **Características Soportadas** | **Características No Soportadas** |
| Creación de Objetos Dinámicos | Carga de clases dinámicas (Dynamic class loading) |
| Paquetes | Garbage Collection and Finalization |
| Métodos virtuales | Threads |
| Herencia | Cloning |
| Interfaces | Palabras claves : *native, synchronized, transient, volatile* |
| Excepciones | Clases no soportadas: *String, Thread, Class,* Wrappers (*Integer, Boolean, Double,* etc.) |
| Palabras clave: *abstract, boolean, break, byte, case, catch, class, continue, default, do, else, extends, final, finally, for, goto, if, implements, import, instanceof, int, interface, new, package, private, protected, public, return, short, static, super, switch, this, throw, throws, try, void, while.* | La clase *java.lang.System* es reemplazada por *javacard.framework.JCSystem*, la cual provee una interface para manejar el comportamiento del sistema |
| Tipos de datos: *boolean, byte, short, int* (opcional). Arrays de una dimensión y de tipos de datos primitivos soportados, arreglos de objetos y de otros arreglos | |

Tabla 8. Características del API de Java Card

## 8. CONCLUSIONES

En este artículo se realizó una breve introducción al tema de los protocolos de cero conocimiento y se explicaron las aplicaciones de estos protocolos sobre implementaciones de esquemas de identificación como residuos cuadráticos, Fiat-Shamir, Guillou-Quisquater y Schnorr además se efectuaron una serie de pruebas para determinar el comportamiento de dichos protocolos sobre un ambiente de tarjetas inteligentes de acuerdo a los criterios de ancho de banda, cálculos realizados, consumo de memoria y tiempo estimado de ejecución, los cuales fueron establecidos previamente para la evaluación realizada.

De acuerdo a los requerimientos y las necesidades de alguna aplicación específica se pueden utilizar diferentes protocolos, para nuestro análisis tuvimos en cuenta básicamente el papel del probador (prover) en un protocolo de cero conocimiento, pues es el papel desempeñado por una tarjeta inteligente, en donde se cuenta con recursos computacionales limitados.

Una vez realizado el análisis, las pruebas y la evaluación de desempeño se concluyo que el protocolo mas indicado para implementar sobre una arquitectura de tarjetas inteligentes es el protocolo de Schnorr pues no requiere realizar muchos cálculos del lado de la tarjeta (requiere básicamente una sola multiplicación modular cuando se esta en línea con el verificador) y simplifica mucho el intercambio de datos entre la tarjeta y su verificador (Servidor, computador, sistema de identificación, entre otros), reduciendo así el consumo de memoria en la tarjeta y facilitando el proceso de identificación para el probador (prover), obviamente agregando complejidad adicional del lado del verificador en comparación con los otros protocolos de cero conocimiento analizados, en particular Fiat-Shamir y Guillou-Quisquater.

Debe tenerse en cuenta que el protocolo de Guillou-Quisquater, permite configurar parámetros como la longitud del vector de inicialización (parámetro $k$), lo cual ahorra el consumo de memoria y el ancho de banda de transmisión (parámetro $t$), esto permite disminuir los datos intercambiados por el probador y el verificador, logrando así adaptarse a las necesidades computacionales y a una arquitectura limitada como la de las tarjetas inteligentes y aplicaciones embebidas pero con su efecto en la seguridad del protocolo, ya que al disminuir estos parámetros se corre un mayor riesgo de aumentar la probabilidad de engaño, lo que debe ser solucionado por el exponente publico $v$, que se utiliza en el protocolo GQ, este debe ser un numero primo grande para garantizar que la probabilidad de un engaño exitoso sea $v^{-kt}$. Cabe anotar que este tipo de reducciones o reconfiguraciones no son posibles en los protocolos de Residuos cuadráticos y Fiat – Shamir, en este ultimo la probabilidad de engaño es $2^{-kt}$, donde $k$ es el numero de secretos de usuario y el parámetro $t$ indica el numero de iteraciones del protocolo para ajustar su nivel de seguridad.

La seguridad de todos estos protocolos de cero conocimiento se basan en la seguridad de los problemas matemáticos sobre los cuales subyacen, por ejemplo, la factorización de un entero compuesto $n$ (RSA); en el protocolo de residuos cuadráticos la dificultad de extraer una raíz cuadrada modulo un entero compuesto $n$; en Fiat-Shamir la dificultad de extraer raíces cuadradas modulo un entero compuesto RSA $n$; en Guillou-Quisquater la dificultad de extraer i-esimas raíces modulo un entero compuesto RSA $n$; en el protocolo de Schnorr la dificultad para calcular logaritmos discretos modulo un numero primo $p$. En la medida en que estos problemas matemáticos se consideren complejos de resolver en términos de recursos, tiempo y dinero, se puede garantizar de algún modo la seguridad de estos protocolos.